\documentclass[multphys,vecphys]{svmult}

\usepackage{makeidx}         
\usepackage{graphicx}        
\usepackage{multicol}        
\usepackage[bottom]{footmisc}

\makeindex             

\newcommand{\gtrsim}{\ ^{\displaystyle >}_{\displaystyle \sim}\ }
\newcommand{\lesssim}{\ ^{\displaystyle <}_{\displaystyle \sim}\ }
\newcommand{\apj}{Astrophys. J.~}
\newcommand{\apjl}{Astrophys. J.~}
\newcommand{\mnras}{Mon. Not. R. Astron. Soc.~}
\newcommand{\aap}{Astron. Astrophys.~}
\begin{document}

\title*{The Difficulty of the Heating of Cluster Cooling Flows by Sound
Waves and Weak Shocks}
\titlerunning{The Difficulty of Heating} 
\author{Yutaka Fujita\inst{1}\and
Takeru Ken Suzuki\inst{2}}
\institute{Department of Earth and Space Science, Graduate School of
Science, \\
Osaka University, Toyonaka, Osaka 560-0043, Japan
\texttt{fujita@vega.ess.sci.osaka-u.ac.jp}
\and Graduate School of Arts and Sciences, University of Tokyo, Komaba,
Meguro, Tokyo 153-8902, Japan \\
\texttt{stakeru@provence.c.u-tokyo.ac.jp}}
%
%
\maketitle

 We investigate heating of the cool core of a galaxy cluster through the
 dissipation of sound waves and weak shocks excited by the activities of
 the central active galactic nucleus (AGN). Using a weak shock theory,
 we show that this heating mechanism alone cannot reproduce observed
 temperature and density profiles of a cluster, because the dissipation
 length of the waves is much smaller than the size of the core and thus
 the wave energy is not distributed to the whole core.

\section{Introduction}
\label{sec:intro}

The failure of standard cooling flow models indicates that the gas is
prevented from cooling by some heating sources. At present, the most
popular candidate for the heating source is the active galactic nucleus
(AGN) at the cluster center. However, it is not understood how the
energy ejected by the AGN is transfered into the surrounding ICM. One
idea is that bubbles inflated by AGN jets move outward in a cluster by
buoyancy and mix the surrounding ICM \cite{chu01,qui01,sax01}. As a
result of the mixing, hot ICM in the outer region of the cluster is
brought into, and subsequently heats, the cluster center. The other idea
is that the dissipation of sound waves created through the AGN
activities. In fact, sound waves or weak shocks that may have evolved
from sound waves are observed in the Perseus and the Virgo clusters
\cite{fab03a,for05}. It was argued that the viscous dissipation of the
sound waves is responsible for the heating of a cool core
\cite{fab03a,fab05}. They estimated the dissipation rate assuming that
the waves are linear. However, when the amplitude of sound waves is
large, the waves rapidly evolve into non-linear weak shocks \cite{ss72},
and their dissipation can be faster than the viscous dissipation of
linear waves. Although it was argued the presence of weak shocks in
Ref.~\cite{fab03a}, their evolution from sound waves was not
considered. Numerical simulations of dissipation of sound waves created
by AGN activities were also performed \cite{rus04}. Their results
actually showed that the sound waves became weak shocks. However, their
simulations were finished before radiative cooling became
effective. Thus, the long-term balance between heating and cooling is
still unknown. In this paper, we consider the evolution of sound waves
to weak shocks, and analytically estimate the `time-averaged' energy
flux of the propagating waves as a function of distance, explicitly
taking account of the dissipation at weak shock fronts and its global
balance with radiative cooling. We assume the Hubble constant of
$H_0=70\:\rm km\: s^{-1}\: Mpc^{-1}$. The detail of the models and
results are shown in Ref.~\cite{fuj05a}.

\section{Models}

We assume that sound waves are created by central AGN activities. The
waves propagate in the ICM outwards. These waves, having a relatively
large but finite amplitude, eventually form shocks to shape sawtooth
waves. If the velocity amplitude is larger than
$\sim 0.1$ sound velocity (the Mach number is $\gtrsim 1.1$), those
waves steepen and become weak shocks after propagating less than a few
wavelengths \cite{suz02}. These shock waves directly heat the
surrounding ICM by dissipating their wave energy. We adopt a heating
model for the solar corona based on a weak shock theory
(\cite{suz02,ss72}, see also \cite{fuj04a}). We assume that a
cluster is spherically symmetric and steady.

The equation of continuity is

\begin{equation}
\label{eq:cont}
 \dot{M}=-4\pi r^2\rho v\:,
\end{equation}
where $\dot{M}$ is the mass accretion rate, $r$ is the distance from the
cluster center, $\rho$ is the ICM density, and $v$ is the ICM
velocity. The equation of momentum conservation is
\begin{equation}
\label{eq:motion}
 v\frac{dv}{dr} = -\frac{GM(r)}{r^2}-\frac{1}{\rho}\frac{dp}{dr}
-\frac{1}{\rho c_s \{1+[(\gamma+1)/2]\alpha_w\}}
\frac{1}{r^2}\frac{d}{dr}(r^2 F_w)
\end{equation}
where $G$ is the gravitational constant, $M(r)$ is the mass within
radius $r$, $p$ is the ICM pressure, $c_s$ is the sound velocity,
$\gamma (=5/3)$ is the adiabatic constant, and $\alpha_w$ is the wave
velocity amplitude normalized by the ambient sound velocity
($\alpha_w=\delta v_w/c_s$). The wave energy flux, $F_w$, is given by
\begin{equation}
\label{eq:Fw}
 F_w = \frac{1}{3}\rho c_s^3 
\alpha_w^2 \left(1+\frac{\gamma+1}{2}\alpha_w\right) \:.
\end{equation}
The energy equation is
\begin{equation}
\label{eq:energy}
\rho v \frac{d}{dr}\left(\frac{1}{2}v^2
+\frac{\gamma}{\gamma-1}\frac{k_B T}{\mu m_H}\right)
+\rho v \frac{G M(r)}{r^2}
+\frac{1}{r^2}\frac{d}{d r}[r^2(F_w+F_c)]
+n_e^2\Lambda(T)=0 \:,
\end{equation}
where $k_B$ is the Boltzmann constant, $T$ is the ICM temperature, $\mu
(=0.6)$ is the mean molecular weight, $m_H$ is the hydrogen mass, $F_c$
is the conductive flux, $n_e$ is the electron number density, and
$\Lambda$ is the cooling function. The term $\nabla\cdot\mbox{\boldmath
$F$}_w$ indicates the heating by the dissipation of the waves. The
equation for the time-averaged amplitude of the shock waves is given by
\begin{equation}
\label{eq:wave}
 \frac{d\alpha_w}{dr}=\frac{\alpha_w}{2}\left[
-\frac{1}{p}\frac{dp}{dr}
-\frac{2(\gamma+1)\alpha_w}{c_s \tau}-\frac{2}{r}
-\frac{1}{c_s}\frac{d c_s}{dr}
\right]\;,
\end{equation}
where $\tau$ is the period of waves, which we assume to be constant
\cite{ss72,suz02}. The second term of the right side of
equation~(\ref{eq:wave}) denotes dissipation at each shock front. 

\section{Results}

For parameters of our model cluster, we adopt the observational data of
the Perseus cluster \cite{ett02}\index{Perseus}. We assume that
$r_s=280$~kpc, $M(r_{1000})=3.39\times 10^{14}\: M_\odot$, and
$r_{1000}=826$~kpc, where the mean density within $r_\Delta$ is $\Delta$
times the critical density of the Universe. Waves are injected at the
inner boundary $r=r_0$, which should be close to the size of bubbles
observed at cluster centers. We assume that $\lambda_0=r_0$, where
$\lambda_0$ is the initial wavelength. If the waves are injected in a
form of sound waves with amplitude $0.1\lesssim\alpha_w < 1$, waves
travel about $\lambda_0$ before they become shock waves
\cite{suz02}. Therefore, for $r_0\leq r \leq r_0+\lambda_0=2\lambda_0$,
we assume that $\nabla \cdot \mbox{\boldmath $F$}_w=0$
(eqs.~[\ref{eq:motion}] and~[\ref{eq:energy}]), and that the second term
of the right-hand side of equation~(\ref{eq:wave}) is zero. The
temperature, electron density, and wave amplitude at $r=r_0$ are $T_0$
$n_{e0}$, and $\alpha_{w0}$, respectively. Unless otherwise mentioned,
the first two are fixed at $T_0=3$~keV and $n_{e0}=0.08\:\rm cm^{-3}$,
respectively, based on the observational results of the Perseus cluster
\cite{san04}.

\begin{figure}
\centering
\includegraphics[height=4cm]{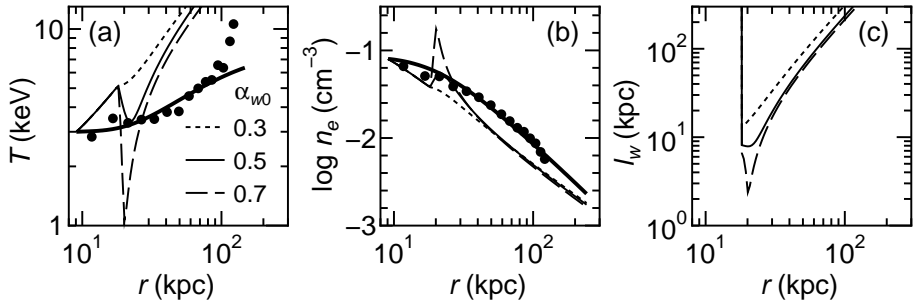}
\caption{(a) Temperature, (b) density,
 and (c) dissipation length profiles for $\alpha_{w0}=0.3$ (dotted lines),
 0.5 (thin solid lines), and 0.7 (dashed lines). Other parameters are
 $\tau=1\times 10^7$~yr, $\dot{M}=50\: M_\odot\:\rm yr^{-1}$, and
 $f_c=0$. Filled circles are {\it Chandra} observations of the Perseus
 cluster \cite{san04}. The bold solid lines correspond to a genuine
 cooling flow model of $\dot{M}=500\: M_\odot\:\rm
 yr^{-1}$.}
\label{fig:alpha}      
\end{figure}

In Figure~\ref{fig:alpha}, we show the results when $\tau$ is fixed at
$1\times 10^7$~yr, and $\alpha_{w0}$ is changed. For the Perseus
cluster, it was estimated that $\alpha_{w0}\sim 0.5$~\cite{fab03a}. The
dissipation length is defined as $l_w=|F_w/\nabla \cdot \mbox{\boldmath
$F$}_w|$. For these parameters, the initial wavelength is
$\lambda_0=9$~kpc, which is roughly consistent with the {\it Chandra}
observations \cite{fab03a}. Other parameters are $\dot{M}=50\:
M_\odot\:\rm yr^{-1}$, and $f_c=0$, where $f_c$ is the ratio of actual
thermal conductivity to the classical Spitzer conductivity.  In general,
larger $\dot{M}$ reproduces observed temperature and density profiles
better. However, large $\dot{M}$ is inconsistent with recent X-ray
observations as was mentioned in \S~\ref{sec:intro}. For comparison, we
show the results of a genuine cooling flow model ($\dot{M}=500\:
M_\odot\:\rm yr^{-1}$, $\alpha_{w0}=0$, and $f_c=0$) and the {\it
Chandra} observations of the Perseus cluster \cite{san04}.
Figures~\ref{fig:alpha}a and~\ref{fig:alpha}b show that only a small
region is heated.  The jumps of $T$ and $n_e$ at $r=2\lambda_0=18$~kpc
are produced by weak shock waves that start to dissipate there. The
energy of the sound waves rapidly dissipates at the shocks, which is
clearly illustrated in short dissipation lengths, $l_w\sim 2$--$15$~kpc
(Fig.~\ref{fig:alpha}c). These dissipation lengths are smaller than
those of viscous dissipation for linear waves, which can be represented
by $l_v=420 \:\lambda_{9}^2\: n_{0.08}\: T_3^{-2}$~kpc, where the
wavelength $\lambda = 9\:\lambda_9$~kpc, the density $n=0.08\:
n_{0.08}\rm\: cm^{-3}$, and the temperature $T=3\: T_3$~keV
\cite{fab03a}. In Figure~\ref{fig:alpha}, the ICM density becomes large
and the temperature becomes small at $r\gtrsim 2\lambda_0$ so that the
rapid shock dissipation is balanced with radiative cooling. Because of
this, waves cannot reproduce the observed temperature and density
profiles that gradually change on a scale of $\sim 100$~kpc.

\begin{figure}
\centering
\includegraphics[height=4cm]{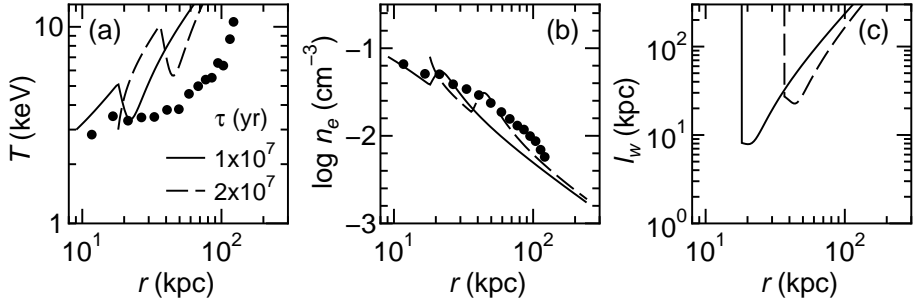}
\caption{Same as Fig.~\ref{fig:alpha}
 but for $\tau=1\times 10^7$~yr (solid lines), and $2\times 10^7$~yr
 (dashed lines). Other parameters are $\alpha_{w0}=0.5$, $\dot{M}=50\:
 M_\odot\:\rm yr^{-1}$, and $f_c=0$. }
\label{fig:tau}      
\end{figure}

In Figure~\ref{fig:tau}, we present the results when $\tau=2\times
10^7$~yr. Compared with the case of $\tau=1\times 10^7$~yr, the wave
energy dissipates in outer regions. However, the dissipation lengths are
still smaller than the cluster core size ($\sim 100$~kpc). Note that
larger $\tau$ (or $\lambda_0$) means formation of larger bubbles. As
indicated in Ref.~\cite{chu00}, it is unlikely that the size of the
bubbles becomes much larger than 20~kpc; the bubbles start rising
through buoyancy before they become larger. On the other hand, when
$\tau<10^7$~yr ($\lambda_0<9$~kpc), the waves heat only the ICM around
the cluster center. The predicted temperature and density profiles are
obviously inconsistent with the observations.

\begin{figure}
\centering
\includegraphics[height=4cm]{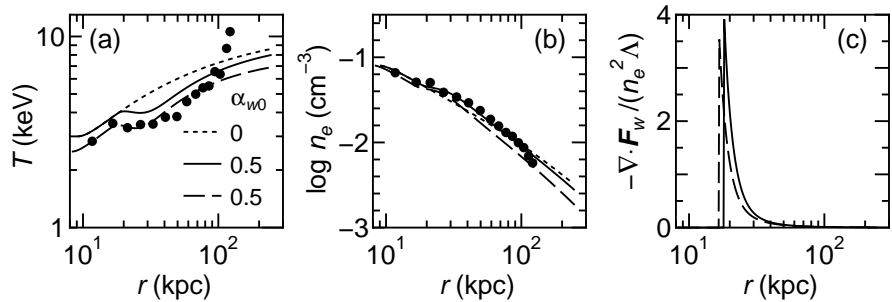}
\caption{(a) Temperature, (b) density,
 and (c) dissipation strength profiles for $\alpha_{w0}=0$ and $T_0=3$~keV
 (dotted lines), $\alpha_{w0}=0.5$ and $T_0=3$~keV (solid lines), and
 $\alpha_{w0}=0.5$ and $T_0=2.5$~keV (dashed lines). Other parameters are
 $\tau=1\times 10^7$~yr, $\dot{M}=50\: M_\odot\:\rm yr^{-1}$, and
 $f_c=0.2$. Filled circles are {\it Chandra} observations of the Perseus
 cluster \cite{san04}.}
\label{fig:fc}      
\end{figure}

The inclusion of thermal conduction changes the situation
dramatically. Figure~\ref{fig:fc} shows the results when $f_c=0.2$. The
models including both wave heating and thermal conduction can well
reproduce the observed temperature and density
profiles. Figure~\ref{fig:fc}c shows the contribution of the wave
heating ($-\nabla \cdot \mbox{\boldmath $F$}_w$) to compensating
radiative cooling ($n_e^2 \Lambda$). Since $-\nabla \cdot
\mbox{\boldmath $F$}_w/n_e^2 \Lambda > 1/2$ for $r\sim 20$--30~kpc, the
wave heating is more effective than the thermal conduction in that
region.

\section{Discussion}

We showed that sound waves created by the central AGN alone cannot
reproduce the observed temperature and density profiles of a cluster,
because the dissipation length of the waves is much smaller than the
size of a cluster core and the waves cannot heat the whole core. The
results have been confirmed by numerical simulations \cite{mat06}. On
the other hand, we found that if we include thermal conduction from the
hot outer layer of a cluster with the conductivity of 20\% of the
Spitzer value, the observed temperature and density profiles can be
reproduced. The idea of the ``double heating'' (AGN plus thermal
conduction) was proposed in~Ref.~\cite{rus02}.

However, the fine structures observed in cluster cores may show that the
actual conductivity is much smaller than that we assumed \cite{fuj02};
the structures would soon be erased, if the conductivity is that
large. If the conductivity is small, we need to consider other
possibilities. While we considered successive minor AGN activities, some
authors consider that rare major AGN activities should be responsible
for heating of cool cores \cite{sok01}. In this scenario, powerful
bursts of the central AGN excite strong shocks and heat the surrounding
gas in the inner region of a cluster on a timescale of $\gtrsim
10^{9}$~yr. Moreover, in this scenario, heating and cooling are not
necessarily balanced at a given time, although they must be balanced on
a very long-term average. This is consistent with the fact that there is
no correlation between the masses of black holes in the central AGNs and
the X-ray luminosities of the central regions of the clusters
\cite{fuj04b}. Alternative idea is that cluster mergers are responsible
for heating of cool cores \cite{fuj04,fuj05}. In this ``tsunami'' model,
bulk gas motions excited by cluster mergers produce turbulence in and
around a core, because the cooling heavy core cannot be moved by the
bulk gas motions, and the resultant relative gas motion between the core
and the surrounding gas induces hydrodynamic instabilities. The core is
heated by turbulent diffusion from the hot outer region of the
cluster. Since the turbulence is produced and the heating is effective
only when the core is cooling and dense, fine-tuning of balance between
cooling and heating is alleviated for this model.



\printindex
\end{document}